\documentclass[12pt]{article}
\usepackage[utf8]{inputenc}
\usepackage{array, tabularx, booktabs}
\usepackage{amsmath,braket}
\usepackage{xcolor}
\usepackage[english]{babel}
\usepackage{graphicx}
\usepackage{jheppub}

\title{Islands in Closed and Open Universes}
\author[a,b]{Raphael Bousso}
\author[a]{and Elizabeth Wildenhain}
\affiliation[a]{Center for Theoretical Physics and Department of Physics,\\
University of California, Berkeley, CA 94720, U.S.A.}
\affiliation[b]{Lawrence Berkeley National Laboratory, Berkeley, CA 94720, U.S.A.}
\emailAdd{bousso@berkeley.edu}
\emailAdd{elizabeth\_wildenhain@berkeley.edu}

\abstract{We show that spatial curvature has a significant effect on the existence of entanglement islands in cosmology. We consider a homogeneous, isotropic universe with thermal radiation purified by a reference spacetime. Arbitrarily small positive curvature guarantees that the entire universe is an island. Proper subsets of the time-symmetric slice of a closed or open universe can be islands, but only if the cosmological constant is negative and sufficiently large in magnitude.}

\begin{document}

\maketitle

\section{Introduction}

Consider the cloud of Hawking radiation resulting from the complete evaporation of black hole formed from collapse.  If the initial and final quantum states are related by a unitary scattering matrix, then the von Neumann entropy $S(t)$ of the Hawking radiation emitted by the intermediate time $t$ should follow the Page curve. That is, $S(t)$ must be the smaller of the coarse-grained radiation entropy and the Bekenstein-Hawking entropy of the remaining black hole at the time $t$. The Quantum Extremal Surface (QES) prescription for computing $S(t)$~\cite{Ryu:2006bv, RyuTak06b, Hubeny:2007xt, Faulkner:2013ana, Engelhardt:2014gca} reproduces this result~\cite{Penington:2019npb,Almheiri:2019psf}. The QES formula can be viewed as an application of the gravitational path integral, in a saddlepoint approximation. Thus, the Page curve---and thus, in particular, the unitarity of the scattering process---can be derived from semiclassical gravity.

The key insight enabling this breakthrough was the recognition that an {\em entanglement island} contributes to the QES formula after the Page time, \emph{i.e.}, during the era when the coarse-grained radiation entropy exceeds the black hole entropy. An island is a portion of the semiclassical spacetime that is topologically disconnected from the reference system or boundary region whose entropy is being computed.

The QES derivation of the Page curve comes on the heels of significant indirect evidence for the unitarity of black hole evaporation, most prominently via the AdS/CFT duality~\cite{Maldacena:1997re}. But the QES formula manages to capture a highly nontrivial aspect of quantum gravity just from a semiclassical analysis. It is vital, therefore, to study its implications in cosmological spacetimes, where we have no other handle on quantum gravity. 

A first objective is to understand whether islands can appear in cosmology. For an evaporating black hole, islands appear naturally when the QES formula is applied to the Hawking radiation. In cosmology, however, it is not clear a priori what process or setup should be considered: what would give rise to the large amounts of entanglement necessary for the formation of an island? What is the relevant reference system (the analogue of the Hawking radiation)? 

One approach to this problem is not to require a natural dynamical origin for the entanglement. Instead, one can consider a simple cosmological solution and make assumptions about the entanglement structure that favor the existence of islands. If islands are absent even under favorable assumptions, this already constitutes an interesting finding. 

In this spirit, Hartman {\em et al.}~\cite{Hartman:2020khs} searched for islands in a radiation-dominated, spatially flat Friedman-Robertson-Walker (FRW) spacetime $M$, entangled with a second nongravitating reference spacetime $M_R$, in a thermofield-double-like state. Instead of first specifying a reference system analogous to the Hawking radiation, Ref.~\cite{Hartman:2020khs} specified spherically symmetric candidate regions $I\subset \Sigma_M$ and asked whether there {\em exists} a reference region $R\subset \Sigma_R$ such that $I$ is an island of $R$. (See \cite{Manu:2020tty,Choudhury:2020hil, Balasubramanian:2020coy} for other work on cosmological islands and thermofield-doubled universes.)

Hartman {\em et al.}\ found that no islands exist unless the cosmological constant is negative, $\Lambda<0$. A flat FRW universe with $\Lambda<0$ expands and then collapses, on a characteristic time scale of order $t_\Lambda \sim |\Lambda|^{-1/2}$. Islands are located in a narrow time band, of order the thermal timescale $\beta\ll t_\Lambda$, before and after the turnaround time; and they must be very large, with proper radius $\gg t_\Lambda$.

\paragraph{Outline and Summary}

Ref.~\cite{Hartman:2020khs} considered only spatially flat FRW universes. In this paper, we will relax this assumption and search for islands in spatially closed and open FRW cosmologies. We will show that a small amount of spatial curvature can have a significant effect. Arbitrarily small positive curvature guarantees that the entire spacetime is an island. It also allows for a new class of islands consisting of more than half (but not all) of the universe. A small---but not arbitrarily small---amount of negative curvature eliminates cosmological islands entirely at fixed $\Lambda<0$. Our results are summarized in Table~\ref{table:summary}.
\begin{center}
\begin{table}[t]
    \centering
    \caption{Summary of Results}
    \label{table:summary}
    \begin{tabular}{p{2.4cm}   p{9.4cm}}
        \toprule
    \textbf{Case} & \textbf{Island Location(s)} \\\midrule
    closed, $\Lambda >0$ & $I=M$\\\midrule
    closed, $\Lambda = 0$ & $I=M$  \\\midrule
    closed, $\Lambda<0$ & $I=M$; and if $t_\Lambda/t_C\lesssim \left(l_P/t_C\right)^{1/2}\ll 1$, then also $I\subsetneq\Sigma_M$, with comoving radius $\chi\in(\chi_*,\pi-\chi_*)$, near turnaround\\\midrule
    open, $\Lambda\geq0$ & None\\\midrule
    open, $\Lambda <0$ & $I\subsetneq\Sigma_M$, with $\chi>\chi^*$, near turnaround, if $t_\Lambda/t_C\lesssim\left(l_P/t_C\right)^{1/2}\ll 1$\\\bottomrule
    \end{tabular}
\end{table}
\end{center}

In Sec.~\ref{prelim} we briefly review the QES prescription and its special case, the island formula. We discuss three necessary conditions that an island $I$ must satisfy~\cite{Hartman:2020khs} regardless of the reference system $R$: $S(I)>A(\partial I)/4G_N$; $I$ is quantum normal; and $G$ is quantum normal, where $G$ is the complement of $I$ on a Cauchy slice $\Sigma_M$ of $M$. We derive a fourth necessary condition that applies only if $M$ is closed and $G$ is nonempty: $S(G)>A(\partial I)/4G_N$. Next, we introduce the specific setting we will study: a spatially closed or open FRW universes with a cosmological constant and radiation. The radiation is entangled with and purified by radiation in an analogous reference spacetime, in a TFD-like state. Finally, we discuss the mode in which our results will be presented: for each class of universe, a physically intuitive analysis of island candidates on the time-reflection symmetric Cauchy slice of $M$ (if present) is followed by a graphical presentation of the validity of the four conditions in the full spacetime solution. 

In Sec.~\ref{closed}, we search for islands in closed FRW solutions (positive spatial curvature). In Sec.~\ref{closedZeroCC}, we consider the simplest case where the cosmological constant $\Lambda$ vanishes. We find that the conditions 1 and 4 discussed in Sec.~\ref{prelim} are mutually exclusive at the turnaround time, so no proper subset of a time-symmetric slice of $M$ is a viable island candidate. This conclusion persists when we analyze the full spacetime. However, we find that $M$ itself is an island, if $R$ contains more than half of $\Sigma_R$.

\begin{figure*}
    \centering
    \includegraphics[width=.8\textwidth]{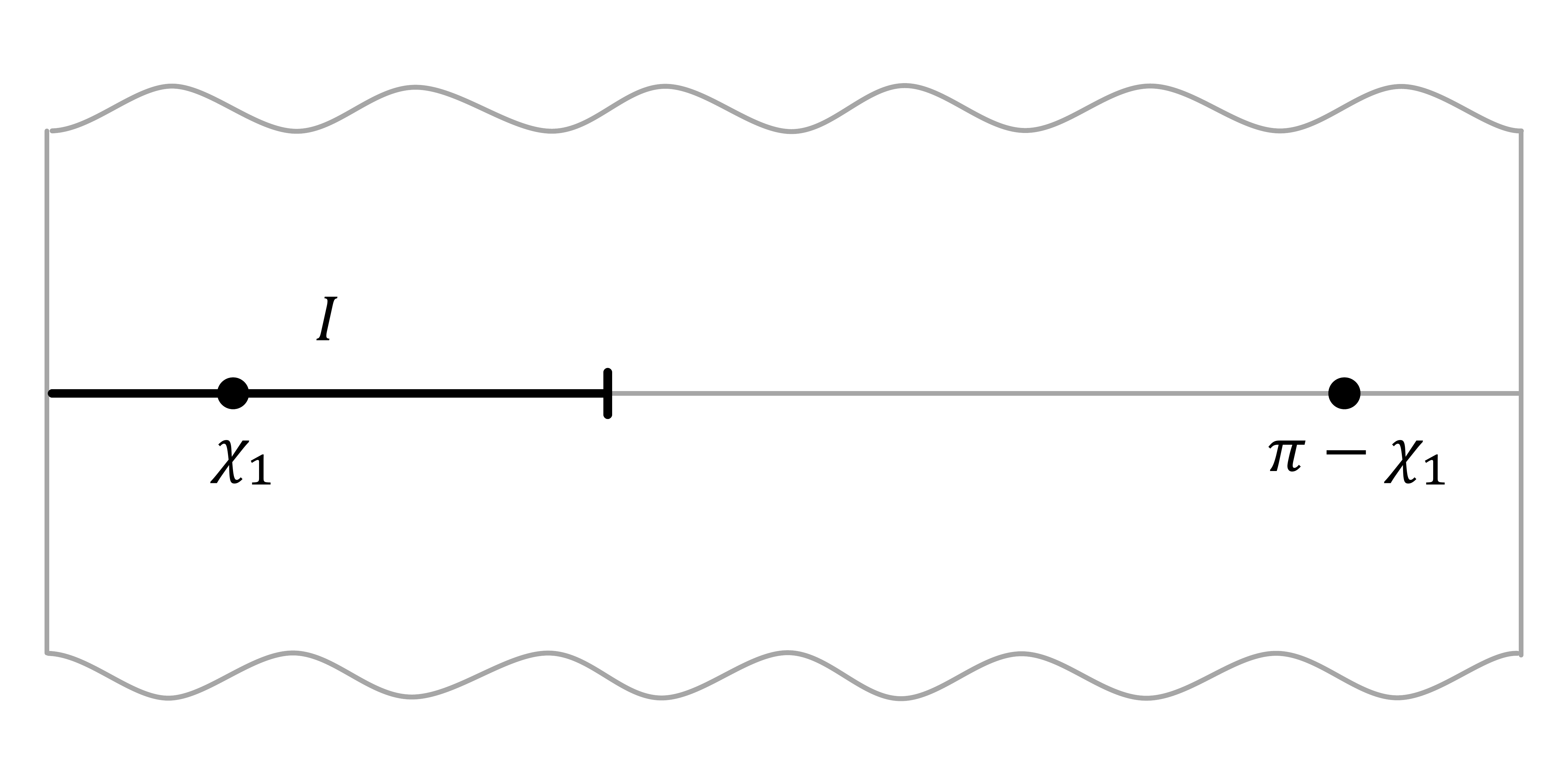}
    \caption{Penrose diagram of a closed recollapsing universe. The entire universe can always be an island. For a proper subset $I$ to be an island, it must lie near the turnaround slice, with boundary within a certain angular range. This range is nonvanishing only if the cosmological constant is negative and sufficiently large.}
    \label{fig:fig1}
\end{figure*}
In Sec.~\ref{closedNegativeCC}, we consider closed universes with negative cosmological constant. The entire universe $M$ is again found to be an island for sufficiently large $R$. For a proper subset of a time-symmetric slice of $M$, we find that the four conditions can be simultaneously satisfied only if the spatial curvature is sufficiently weak (and dynamically irrelevant) at the turnaround time. A check of the full solution indicates that islands only appear near the turnaround time. In this case we find explicit examples of islands that are a proper subset of the time-symmetric slice of $M$, with $R$ being the analogous region on $\Sigma_R$; see Fig.~\ref{fig:fig1}. We also find examples of regions that satisfy all three necessary conditions of Ref.~\cite{Hartman:2020khs} but which are not islands for any choice of $R$, because they fail to satisfy the condition 4.

In Sec.~\ref{closedPositiveCC}, we examine closed universes with positive cosmological constant. $M$ itself is again an island if $R$ contains more than half of $\Sigma_R$. We find that no proper subset of the time-symmetric Cauchy slice of $M$ can be an island, as conditions 1 and 4 are mutually incompatible. A check of the full solution confirms that no proper subset of any other Cauchy slice can be an island.

\begin{figure*}
    \centering
    \includegraphics[width=.8\textwidth]{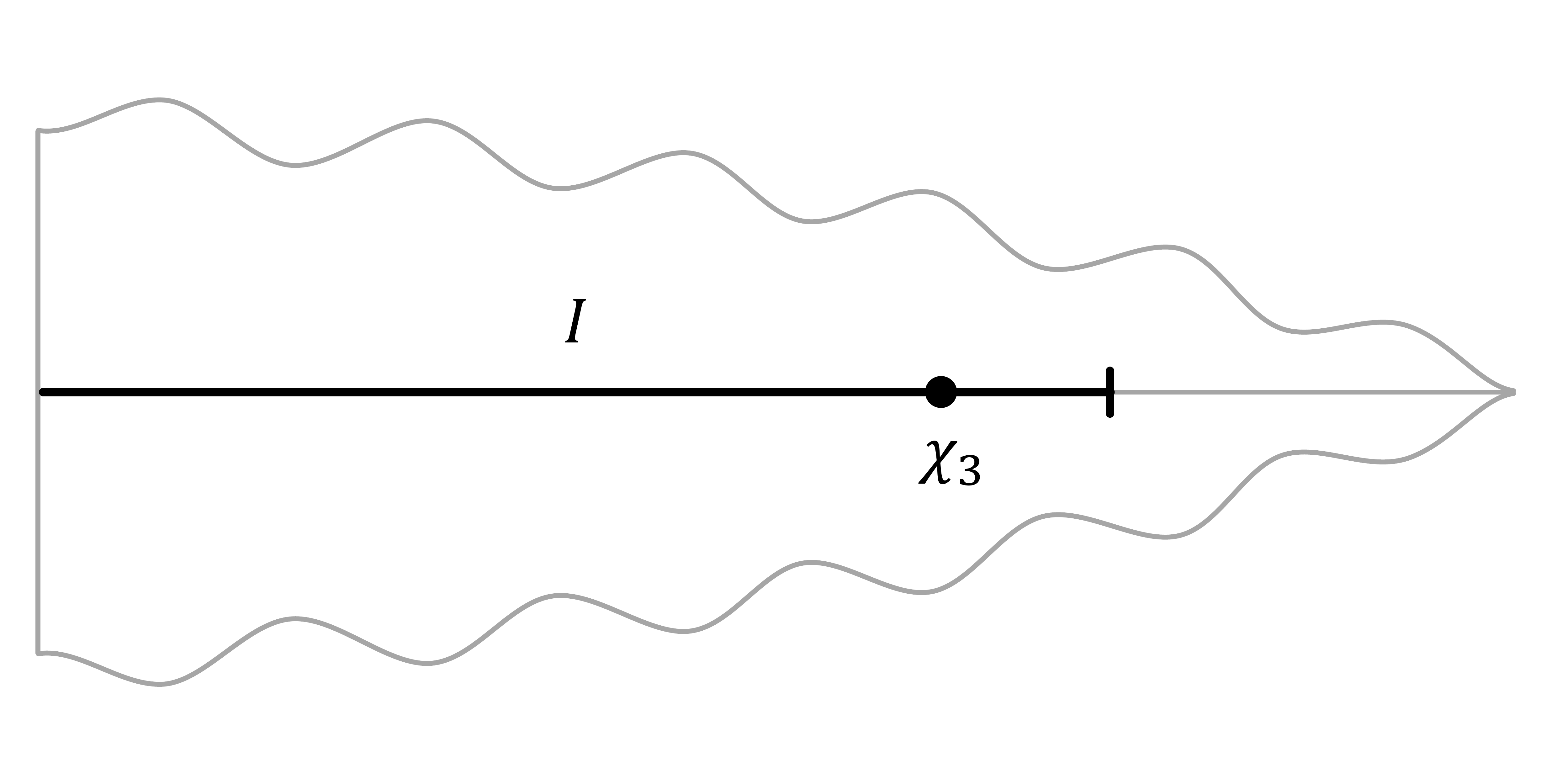}
    \caption{Penrose diagram of an open universe. A sufficiently large region $I$ at the turnaround time is an island, if the cosmological constant is negative and large enough for curvature not to dominate below the critical radius $\chi_3$.}
    \label{fig:fig2}
\end{figure*}

In Sec.~\ref{open}, we turn to solutions with negative spatial curvature (open FRW). If $\Lambda\geq 0$, there are no islands. If $\Lambda<0$, we find islands exist if the spatial curvature radius is at least comparable to the minimum island size in the spatially flat case; see Fig.~\ref{fig:fig2}. This is easy to understand geometrically: for $I$ to be an island, one must have $S(I)>A(\partial I)/4 G_N$. The entropy is extensive. In flat space, volume grows faster than area, so this condition becomes satisfied at large radius. But in a hyperbolic geometry, volume and area approach a fixed ratio for radii greater than the curvature radius. Therefore, the condition does not become automatically satisfied for sufficiently large radius.

\paragraph{Relevance to the Observed Universe} Our primary motivation is to examine whether islands can exist in a larger class of cosmological models, regardless of whether they describe our own universe. But it is interesting to ask whether we could live on an island. Let us briefly discuss the extent to which our analysis constrains this question.

Like Ref.~\cite{Hartman:2020khs}, we consider only universes with radiation and a cosmological constant $\Lambda$ of arbitrary sign. The visible universe has $\Lambda>0$~\cite{SupernovaCosmologyProject:1998vns,Riess:1998}, so only the $\Lambda> 0$ case is directly relevant to the question of whether \emph{we} live on an island. However, in a theory with multiple vacua, the universe may decay to regions with $\Lambda=0$ or $\Lambda<0$ in the future. Thus, results for $\Lambda \leq 0$ could still be relevant for future regions in our own universe.

The observed universe went through a long era between radiation and vacuum domination, when the pressure was negligible. The matter-dominated era is not explicitly modeled here. But since matter and radiation both have extensive entropy, we do not expect qualitatively different results. We anticipate a reduction in the allowed size range for matter islands compared to radiation islands.

The visible universe is consistent with exact spatial flatness~\cite{Planck:2018vyg}. However, observational constraints only put a lower bound on the curvature radius; they do not rule out spatial curvature on a scale somewhat larger than the visible universe. This is significant, since we show that an arbitrarily small (hence locally unobservable) amount of positive curvature allows for islands in a universe with $\Lambda>0$.

Indeed, exact spatial flatness requires infinite fine-tuning. Moreover, approximate spatial flatness is a dynamical repeller during matter and radiation domination. Hence, even the approximate spatial flatness of the visible universe dictated by observational constraints would require tremendous fine-tuning of initial conditions, unless it is the result of a dynamical process such as slow-roll inflation. (We are not aware of any other viable candidate process.) This is a period of accelerated expansion driven, for example, by a scalar field with a slowly varying positive potential~\cite{Linde:1983gd}. Inflation will not make space exactly flat. 

The visible universe is anisotropic and inhomogeneous at scales below 100 Mpc, whereas the FRW approximation we use assumes exact spatial homogeneity and isotropy. However, this should not affect any conclusions about cosmological islands. A necessary condition for $I$ to be an island is that its matter entropy must exceed the Bekenstein-Hawking entropy of its boundary $\partial I$. This can only happen for regions much larger than the horizon scale, for which the FRW description is a good approximation. (Of course, black hole islands could exist if a small black hole forms and evaporates.)
 
A failure of homogeneity and isotropy at scales much larger than the visible universe is expected in plausible cosmological models~\cite{Bousso:2000xa}. This may lead to additional classes of islands. For example, if our universe descended from a metastable vacuum with larger $\Lambda$, its homogeneity and isotropy on slices of constant density is a consequence of the symmetries of the dominant instanton mediating false vacuum decay~\cite{Coleman:1980aw}. But on scales that include the parent vacuum and other baby universes, the spacetime admits no preferred slicing on which it would appear homogeneous; it is not an FRW solution. In a complex multiverse, such as the spacetime that would arise in the landscape of string theory, even the number of noncompact dimensions could change over large scales. It is interesting to ask whether there are new classes of islands in such models, especially islands associated to ``hat'' regions with $\Lambda=0$~\cite{Bousso:2011up, Susskind:2007pv, Aguirre:2010rw, Nomura:2011dt}. This is an interesting possibility~\cite{Langhoff:2021uct}, whose general study we leave to future work.
   
\section{Preliminaries}
\label{prelim}

\subsection{Quantum Extremal Surface Prescription}

The all-orders~\cite{Engelhardt:2014gca} quantum-corrected~\cite{Faulkner:2013ana}, covariant~\cite{Hubeny:2007xt} Ryu-Takayanagi~\cite{Ryu:2006bv,RyuTak06b} prescription (QES prescription) computes the entropy of a nongravitating system $R$ in terms of a dual spacetime with gravity, $M$, whose state and geometry are computed semiclassically:
\begin{equation}
    S(\mathbf{R}) = S_{\rm gen}[\mathrm{EW}(R)]~.
    \label{srer}
\end{equation}
The bold-face notation~\cite{Almheiri:2019yqk} distinguishes the (presumably correct) entropy computed by the QES formula from the von Neumann entropy $S(R)$ computed directly from the semiclassical state. Here $\mathrm{EW}(R)$ is the entanglement wedge and $S_{\rm gen}$ is its generalized entropy. We will now briefly summarize their definitions. 

For a partial Cauchy surface $X\subset \Sigma_M$,
\begin{equation}
S_{\rm gen}(X) = \frac{\text{Area}[\partial X]}{4G_N\hbar} + S(X)~,
\end{equation}
where $S(X)$ is the von Neumann entropy of the density operator of the quantum field theory state reduced to $X$. Both terms are cutoff-dependent, but their sum is well-defined (see the appendix in Ref.~\cite{Bousso:2015mna} for a detailed discussion).

$\mathrm{EW}(R)$ is a spacetime region whose generalized entropy is ``extremal'' (really, stationary) with respect to small shape deformations of its boundary surface in $M$, subject to certain homology and global minimality conditions. (Settings with highly incompressible quantum states require a more precise definition~\cite{Akers:2020pmf}. In doubly holographic settings, the appropriate homology rule must be chosen with care~\cite{Bousso:2020kmy}. Neither subtlety will arise here. The entanglement wedge is in general state-dependent; it is related through a choice of code subspace to the reconstructible wedge~\cite{Hayden:2018khn,Akers:2021fut}, which is not. We implicitly assume a small code subspace in this paper, so that we can neglect this distinction.)

Now let us specialize to the case where $R\subset \Sigma_R$ is a partial Cauchy surface in a nongravitating spacetime $M_R$ distinct from $M$. Then the definition of $\mathrm{EW}(R)$ reduces to the ``island rule''~\cite{AMMZ}:
\begin{enumerate}
    \item $\mathrm{EW}(R)=I\cup R$, where $I\subset \Sigma_M$ and $I$ is compact;\footnote{More precisely, the homology rule requires that in the conformally compactified spacetime, the boundary of the image of $I$ does not intersect with the conformal boundary of $M$.} 
    \item $S_{\rm gen}(I\cup R)$ is stationary under any local variations of the boundary surface $\partial I$; 
    \item Among all such regions globally, $I$ yields the smallest $S_{\rm gen}(I\cup R)$.
\end{enumerate}
Note that $I=\varnothing$ is allowed.

For example, suppose that $M_R$ is coupled to $M$, and that $R$ contains the Hawking radiation emitted by an evaporating black hole prior to the time $t$~\cite{Penington:2019npb, Almheiri:2019psf}. ($R$ could also be a weakly gravitating distant region containing the radiation~\cite{Bousso:2019ykv}.) In the semiclassical approximation, the radiation is thermal~\cite{Hawking:1974sw}. Its entropy $S(R(t))$ increases monotonically, implying information loss~\cite{Hawking:1976ra}. However, after the Page time, the entanglement wedge $\mathrm{EW}(R)$ includes an island $I\neq\varnothing$ that purifies the radiation~\cite{Penington:2019npb, Almheiri:2019psf}. The island is the black hole interior slightly before the most recent radiation in $R$ was emitted. Thus, $S_{\rm gen}[\mathrm{EW}(R)]$ is dominated by the area term $A[\partial I]/4G_N$, which decreases as the black hole shrinks.

\subsection{Four Necessary Conditions for Islands}
\label{sec-necessary}

A nonempty island $I$ must satisfy four conditions that do not depend on $R$. We will begin by deriving the first three, following Ref.~\cite{Hartman:2020khs}. Since $I\neq \varnothing$, we have
\begin{equation}
    S(R)>S_{\rm gen}(I\cup R)=\frac{A(\partial I)}{4G_N} + S(I\cup R)\geq 
    \frac{A(\partial I)}{4G_N} + S(R)-S(I)
\end{equation}
by subadditivity of the von Neumann entropy; hence
\begin{equation}
    \label{equation:condition1}
    S(I)>\frac{A(\partial I)}{4G_N}~.~~~~\mbox{(Condition 1)}
\end{equation}

By assumption, $I\cup R$ is quantum extremal, \emph{i.e.}, $S_{\rm gen}(I\cup R)$ is stationary under shape deformations of $\partial I$. The area contribution to this variation does not change when we consider $S_{\rm gen}(I)$ instead; and by strong subadditivity, the shape derivative of the von Neumann entropy in the past or future directions outward from $I$ can only increase when $R$ is dropped~\cite{Bousso:2015mna}. Hence it must be non-negative:
\begin{equation}
\label{equation:condition2}
I~\mbox{is quantum normal.~~~~(Condition 2)}
\end{equation}

We will take the global quantum state on $\Sigma_M\cup \Sigma_R$ to be pure. (This can always be arranged by adding a purifying auxiliary system to $M_R$.) Hence $G\cup Q$ is also quantum extremal, where $G\equiv \Sigma_M \backslash I$ and $Q\equiv \Sigma_R\backslash R$. The above argument implies that
\begin{equation}
\label{equation:condition3}
G~\mbox{is quantum normal.~~~~(Condition 3)}
\end{equation}

$M$ always satisfies extremality. However, it satisfies the homology condition only if its Cauchy surfaces are closed. For a proper subset $I\subsetneq \Sigma_M$ to be an island, in this case, it must be a better candidate than the whole of $M$:
\begin{equation}
\label{smrir}
    S_{\rm gen}(M\cup R)>S_{\rm gen}(I\cup R)~.
\end{equation}
Since $M$ is spatially closed, $S_{\rm gen}(M\cup R)=S(M\cup R)$, and Eq.~\eqref{smrir} implies
\begin{equation}
    S(I\cup R) + \frac{A(\partial I)}{4G_N} < S(M\cup R) \leq S(G)+S(I\cup R)
\end{equation}
by subadditivity of the von Neumann entropy. Hence we find a fourth condition:
\begin{equation}
    \label{equation:condition4}
    \mbox{For spatially closed}~M~\mbox{and}~I\subsetneq \Sigma_M~:~~ S(G)>\frac{A(\partial I)}{4G_N}~.~~~~\mbox{(Condition 4)}
\end{equation}

\subsection{Thermofield-doubled FRW Universes}

In the next two sections, we shall search for islands in cosmological spacetimes. We will consider a spatially homogeneous and isotropic universe $M$ in 4 dimensions with positive or negative spatial curvature, thermal radiation, and arbitrary cosmological constant $\Lambda$. The metric is
\begin{equation}
    ds^2 = -dt^2 + a(t)^2\left(d\chi^2+f^2(\chi)d\Omega^2\right)~,
\end{equation}
where $a(t)$ is the scale factor. The function $f(\chi)$ depends on the curvature: $f(\chi) = \sinh(\chi), \chi$, or $\sin(\chi)$ for open, flat, and closed universes respectively. Another convenient coordinate system uses conformal time $\eta$, defined via
\begin{equation}
    d\eta = \frac{dt}{a(t)}~.
\end{equation}
In these coordinates, the FRW metric takes the form
\begin{equation}
    ds^2 = a^2 (\eta)\left(-d\eta^2 + d\chi^2 + f^2(\chi)d\Omega^2\right).
\end{equation}
The scale factor $a(t)$ obeys the Friedmann equation:
\begin{equation}\label{fried1}
    \left(\frac{\dot{a}}{a}\right)^2 = \frac{8\pi G_N \rho_r}{3} + \frac{\Lambda}{3} - \frac{k}{a^2}~,
\end{equation}
where $\rho_r$ is the energy density of radiation and $\Lambda$ is the cosmological constant. 

We will mainly be interested in universes with an initial curvature singularity (a big bang). Solutions with a big crunch but no big bang are trivially related by time-reversal. Radiation redshifts as $\rho_r\propto a^{-4}$, so its energy density will dominate near the big bang, \emph{i.e.}, at sufficiently early times. 

The cosmological constant will come to dominate the evolution within a time of order
\begin{equation}
    t_\Lambda \equiv \sqrt{3/|\Lambda|}
\end{equation}  
after the big bang, if the universe reaches this age.

At the time
\begin{equation}
    \label{equation:tc}
    t_C \equiv \left(\frac{8\pi G_N \rho_r a^4}{3}\right)^{1/2}~,
\end{equation}
after the big bang, the curvature term in the Friedmann equation begins to dominate over the radiation term. If the universe reaches this age, and if $t_C<t_\Lambda$, a curvature-dominated era begins at $t_C$ and ends at $t_\Lambda$. For recollapsing solutions, the same sequence happens in reverse after the turnaround time. 

Solutions without a singularity arise only if the cosmological constant and curvature are both positive and the radiation density at the turnaround time is sufficiently small. Then the above definitions can be still be made, but they do not have the stated physical interpretation. Moreover, $t_C$ and $t_\Lambda$ do not fix a solution uniquely. Hence we will use a different parametrization of solutions in Sec.~\ref{closedPositiveCC}.

It will be convenient to express the Friedmann equation in terms of $t_C$ and $t_\Lambda$:
\begin{equation}\label{fried2}
    \left(\frac{\dot{a}}{a}\right)^2 = \frac{t_C^2}{a^4} \pm \frac{1}{t_\Lambda^2} - \frac{k}{a^2}~. 
\end{equation}
The $\pm$ corresponds to the sign of $\Lambda$. 

Throughout this paper we shall assume that the effective number of light fields is of order unity. (It is easy to generalize to a larger number of fields, and strictly it is necessary to do so in order to justify neglecting the contribution of gravitons to the entropy. But increasing the number of radiation species does not lead to new regimes in our analysis, while it does complicate the formulas.) Then the physical entropy density of the thermal radiation is
\begin{equation}
    \label{equation:sphys}
    s \sim \rho_r^{3/4}~,
\end{equation}
and the comoving entropy density is
\begin{equation}
    \label{equation:s}
    s_c \equiv s a^3 \sim  \left(\frac{t_C}{l_P} \right)^{3/2}~,
\end{equation}
where $l_P\equiv G_N^{1/2}$ is the Planck length.

Following Ref.~\cite{Hartman:2020khs}, we purify the thermal radiation by invoking a second, nongravitating spacetime $M_R$ and constructing a thermofield double. $M_R$ is defined up to conformal transformations; here we choose
\begin{equation}
    ds_R^2/\ell^2 = -d\eta_R^2 + d\chi_R^2 + f^2(\chi_R)d\Omega_R^2~,
\end{equation}
where $\ell$ is an arbitrary fixed length scale. The thermofield double is first constructed using two copies of $M_R$:
\begin{equation}
\ket{\mbox{TFD}} \propto \sum_n e^{-\beta E_n} \ket{n}^*_1 \ket{n}_2~; 
\end{equation}
then a conformal transformation by $a^2$ is applied to transform one copy to $M$. Here $\beta=\ell/(aT)$, where $T$ is the radiation temperature in the physical spacetime $M$ at scale factor $a$. Our convention for the respective time orientations is opposite to that of Ref.~\cite{Hartman:2013qma}; see Refs.~\cite{Hartman:2020khs} for further details.

We note an important property of the thermofield double which will be useful below. For regions $I\subset \Sigma_M$ and $R\subset \Sigma_R$ with equal coordinate position, the renormalized\footnote{The entropy of bounded regions in QFT has universal short-distance divergences that can be stripped off so long as the characteristic wavelength of excitations is greater than the Planck scale. In this paper, we only consider regions that are under semiclassical control, $(a/\ell) \beta\gg l_P$.} entropy of $I\cup R$ is small and increases when $I$ and $R$ are separated in time at fixed comoving size. It also increases if the size of either $I$ or $R$ is increased or decreased at fixed time:
\begin{equation}
\label{sir}
    S(I\cup R)\approx s_c |\Delta V_c|~,
\end{equation}
where $\Delta V_c=V^R_c-V^I_c$, and $V_c$ denotes comoving volume (\emph{i.e.}, $V^R_c=V(R)/\ell^3$ and $V^I_c=V(I)/a^3$). The sharp transition in Eq.~\eqref{sir} when $I$ and $R$ coincide is smoothed on the thermal scale $\beta$~\cite{Hartman:2020khs}.

\subsection{General Analysis and Restriction to Time-Symmetric Slices}

For each class of universes, we will search for spherical\footnote{If $I$ is a spherical island of the region $R$, then one expects that deformations of $R$ will still have an island that is a small deformation of $I$. Area is ``expensive'' so generally $I$ will deviate less from spherical symmetry than $R$. Our analysis does not rule out the existence of different classes of islands that are not approximately spherical.} islands by checking the four necessary conditions laid out in Sec.~\ref{sec-necessary}. This check is performed using solutions for the full spacetime, and the results are displayed in plots showing where each condition is satisfied. 

The solutions for the full spacetime are relatively complicated. In order to develop some intuition, we precede each full analysis by searching for islands only on time-symmetric slices. Ultimately, this where we expect to find islands, because quantum extremality is difficult to satisfy when the universe is expanding or contracting. 


Quantum extremality requires that the classical expansion is compensated by the time-derivative of the renormalized entropy at fixed $\chi$. This is possible if the classical expansion is itself very small, of order $G$, \emph{i.e.}, in a small time interval around the turn-around time. In the spatially flat case, the size of this interval is of order $\beta$~\cite{Hartman:2020khs}. The same conclusion applies to open and closed universes: as we shall see below, curvature is dynamically negligible at the turnaround time in all cases where we find islands at that time.

On time-symmetric slices, the necessary conditions of Sec.~\ref{sec-necessary} take a special form. The scale factor at the turnaround time, $a_0$, is found by setting $\dot a=0$ in Eq.~\eqref{fried2}. Then the conditions become
\begin{equation}
    \label{equation:condition1TS}
    s_c V_c(\chi)\geq\frac{a_0^2 A_c(\chi)}{4l_P^2}~~~~~\mbox{(Condition 1)}
\end{equation}
\begin{equation}
    \label{equation:condition2TS}
    \frac{\partial}{\partial \chi}S_{\mathrm{gen}}[I(\chi)]\geq 0~~~~~\mbox{(Condition 2)}
\end{equation}
\begin{equation}
    \label{equation:condition3TS}
    -\frac{\partial}{\partial \chi}S_{\mathrm{gen}}[G(\chi)]\geq 0~~~~~\mbox{(Condition 3)}
\end{equation}
\begin{equation}
    \label{equation:condition4TS}
    s_c \left(V_c^{\mathrm{tot}} - V_c(\chi)\right)\geq\frac{a_0^2 A_c(\chi)}{4l_P^2}~~~~\mbox{(Condition 4)}
\end{equation}
for a spherical island candidate of radius $\chi$. Here $V_c^{\mathrm{tot}}$ is the comoving volume of the entire closed universe at the turnaround time, and
\begin{equation}
    S_{\mathrm{gen}} = s_c V_c(\chi) + \frac{A(\chi)}{4G_N}~.
\end{equation}

In the next two sections, we will analyze the closed and open cases, respectively. We will examine whether the necessary conditions can be satisfied, and if so, we will check whether they are sufficient.

\section{Closed Universes}
\label{closed}

In this section, we consider solutions with positive spatial curvature (closed FRW). In such a geometry, the coming volume and area functions on the unit three-sphere are
\begin{gather}\label{closedvolume}
    V_c = \pi (2\chi - \sin2\chi)~,\\
    A_c = 4\pi \sin^2\chi~.\label{closedarea}
\end{gather}

\subsection{Positive Curvature, Zero Cosmological Constant}
\label{closedZeroCC}

The first closed universe we consider is the simplest: one with $\Lambda=0$. We begin our search for islands by restricting our attention to the time-symmetric slice. We consider spherically symmetric regions, $I(\chi)$, which extend from the origin to the sphere at $\chi$ at the turnaround time. For $\Lambda=0$ and $k=1$, the scale factor at the turnaround time satisfies
\begin{equation}
    0 = \frac{t_C^2}{a_0^4} - \frac{1}{a_0^2}
\end{equation}
by Eq.~\eqref{fried2}. Hence
\begin{equation}
    a_0 = t_C~.
\end{equation}

\begin{figure*}
    \centering
    \includegraphics[width=.8\textwidth]{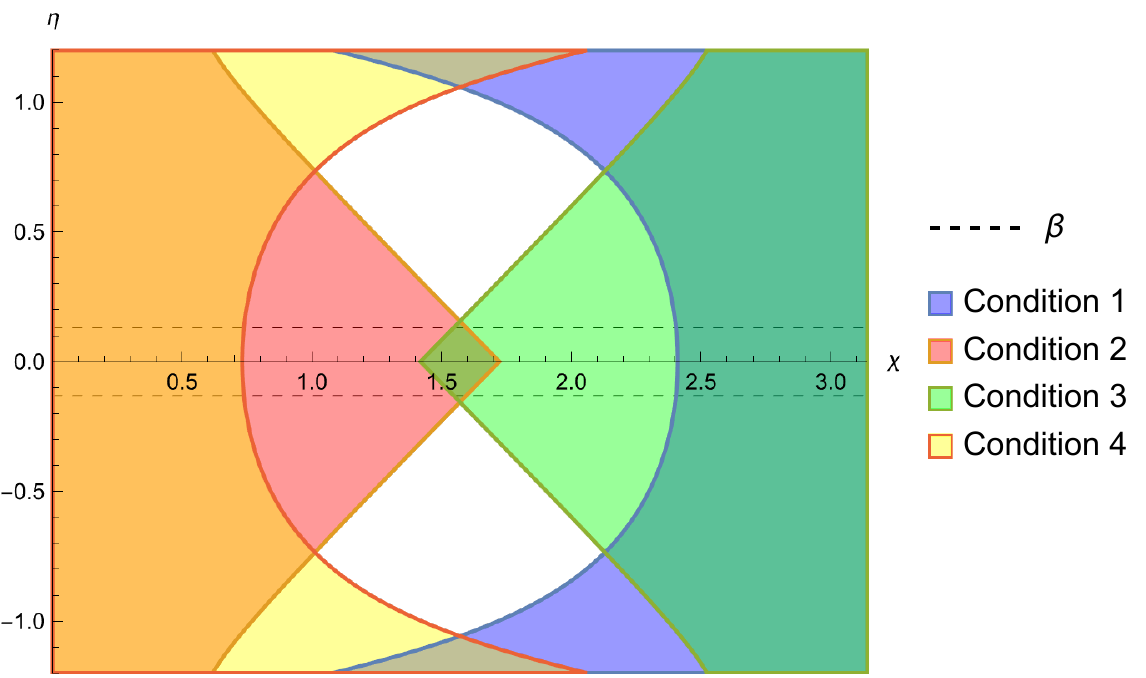}
    \caption{Regions satisfying the four island conditions are shown for a closed universe with $\Lambda=0$. The radiation temperature is $\beta^{-1}$ at the turnaround time $t=\eta=0$. Chosen for display is $t_C=170 t_P$. Top and bottom cutoffs are chosen so as to eliminate artifacts of the Planck regime near the big bang and big crunch. The lack of four-way overlap shows that no region satisfies all four conditions, so there cannot be any islands.}
    \label{fig:plotPK0CC}
\end{figure*}

Since the universe is closed, we can consider either the entire universe $M$, or a proper subset $I(\chi)$, $\chi<\pi$, of its time-symmetric slice, as an island candidate. We begin by ruling out the latter, by showing that conditions 1 and 4 are mutually incompatible. 

Condition 1 is that the radiation entropy in $I$ exceed the Bekenstein-Hawking entropy of the boundary. Using the time-symmetric version of condition 1, Eq.~\eqref{equation:condition1TS}, and Eq.~\eqref{equation:s}, this becomes
\begin{equation}
    \left(\frac{t_C}{l_P}\right)^{3/2} V_c\gtrsim \frac{t_C^2 A_c(\chi)}{l_P^2}~,
\end{equation}
which is equivalent to 
\begin{equation}
    \frac{V_c(\chi)}{A_c(\chi)}\gtrsim \left(\frac{t_C}{l_P}\right)^{1/2}~.
\end{equation}

The ratio of comoving volume to area is of order $\chi$ for small $\chi$ and grows monotonically, diverging as $\chi\to \pi$. We are only interested in the semiclassical regime,
\begin{equation}
    \frac{t_C}{l_P} \gg 1~.
\end{equation}
Thus condition 1 requires $\pi-\chi\ll 1$; that is, the island must be nearly the whole universe. In this regime,
\begin{equation}
    \frac{V_c}{A_c}\approx \frac{2\pi^2}{4\pi(\pi-\chi)^2} = \frac{\pi}{2(\pi-\chi)^2}~.
\end{equation}
and condition 1 becomes
\begin{equation}
    \pi-\chi \lesssim \left(\frac{l_P}{t_C}\right)^{1/4}~,
\end{equation}

Condition 4 requires that the radiation entropy in $G$, the complement of $I$ in $M$, should also exceed the Bekenstein-Hawking entropy of its boundary. Hence $G$ must also consist of nearly the entire time-symmetric slice. But this contradicts the definition of $G$: $G$ and $I$ cannot be mutual complements and both consist of nearly all of $\Sigma_M$. Since the necessary conditions 1 and 4 cannot be simultaneously satisfied, no proper subset of the time-symmetric slice can be an island.

To check that the restriction to the time-symmetric slice did not miss viable island candidates, we examine the full solution. The scale factor is:
\begin{equation}
    a(\eta) = t_C \cos(\eta)~,
\end{equation}
with the turnaround time set at $\eta=0$. Fig.~\ref{fig:plotPK0CC} shows the regions in which the four conditions are satisfied. As expected, there is no region of four-way overlap. Therefore, no proper subset of a closed universe with $\Lambda=0$ can be an island.

Next, we turn to $M$ itself as an island candidate. $M$ trivially satisfies all necessary conditions, since it has no boundary. But for $M$ to be an island, it must beat the empty set; we require $S(R\cup M)<S(R)$. Since $S(R\cup M)\approx s_c [V_c(\Sigma_R)-V_c(R)]$, $M$ is an island of $R$ if and only if $R$ is more than half of $\Sigma_R$.

\subsection{Positive Curvature, Negative Cosmological Constant}
\label{closedNegativeCC}

Now we consider closed universes with $\Lambda <0$. As before, we begin with a restriction to time-symmetric slices. If $t_\Lambda\gg t_C$, then $\Lambda$ is insignificant at the turnaround time. Then $\Lambda$ never plays a dynamical role, and we expect the results to be the same as the $\Lambda = 0$ case. Thus, we will consider the regime 
\begin{equation}\label{tltc}
    t_\Lambda \lesssim t_C~.
\end{equation}
By Eq.~\eqref{fried2}, the scale factor satisfies
\begin{equation}\label{fried3}
    0 = \frac{t_C^2}{a_0^4} -\frac{1}{t_\Lambda^2} - \frac{1}{a_0^2}
\end{equation}
at the turnaround time, so with Eq.~\eqref{tltc} we find
\begin{equation}\label{equation:aNCC}
    a_0\sim \sqrt{t_\Lambda t_C}~.
\end{equation}

Consider first a proper subset of the universe as the island candidate: $I(\chi)$ with $\chi<\pi$. Condition 1 was given in Eq.~\eqref{equation:condition1TS}. Substituting the above expression for $a_0$ we find
\begin{equation}\label{c1closedneg}
    \frac{V_c(\chi)}{A_c(\chi)}> \alpha\sim
    \left(\frac{t_\Lambda}{t_C}\right)\left(\frac{t_C}{l_P}\right)^{1/2}~,
\end{equation}
where we have defined the combination of parameters $\alpha$ for later convenience.

Condition 4 yields the same inequality with $\chi\to\pi-\chi$. Since $V_c/A_c$ is monotonic in $\chi$, conditions 1 and 4 can be satisfied simultaneously only if
\begin{equation}\label{alphalimit}
    \alpha< \frac{V_c(\pi/2)}{A_c(\pi/2)}=\frac{\pi}{4}~.
\end{equation}
Hence we require
\begin{equation}\label{closedc1regime}
    \frac{t_\Lambda}{t_C}\lesssim \left(\frac{l_P}{t_C}\right)^{1/2}\ll 1~,
\end{equation}
where the second inequality is the condition for a classical solution. Note that this conclusion disallows $t_\Lambda\sim t_C$ and hence is stronger than Eq.~\eqref{tltc}.

In the regime characterized by Eq.~\eqref{closedc1regime}, solving Eq.~\eqref{c1closedneg} as an equality yields a critical value $\chi_1<\pi$ such that the inequality \eqref{c1closedneg} will be satisfied for all $\chi>\chi_1$. Hence, conditions 1 and 4 will be simultaneously satisfied for
\begin{equation}\label{c14range}
    \chi_1<\chi< \pi-\chi_1~.
\end{equation}
In the limit as $\alpha\ll 1$, one finds $\chi_1\sim \alpha$.

We turn to conditions 2 and 3. Condition 2 is the requirement that the island be quantum normal. Using the time-symmetric version of condition 2, Eq.~\eqref{equation:condition2TS}, and Eqs.~\eqref{equation:s}, \eqref{closedvolume}, \eqref{closedarea},  and \eqref{equation:aNCC} we find
\begin{equation}\label{cotchi}
    \cot(\chi)\gtrsim - \alpha^{-1}~.
\end{equation}
Note that the cotangent monotonically decreases in the range $\chi\in(0,\pi)$ and becomes negative for $\chi>\pi/2$. In the regime where conditions 1 and 4 can be satisfied, the magnitude of the right hand side is at least of order unity by Eq.~\eqref{alphalimit}. Hence, the above condition corresponds to
\begin{equation}
    0<\chi<\pi-\chi_2~,
\end{equation}
where 
\begin{equation}
    \chi_2<\frac{\pi}{2} ~\mbox{and}~ \frac{\pi}{2}-\chi_2\sim O(1)~.
\end{equation}

Condition 3 mandates that $G$ (the complement of $I$) be quantum normal; by symmetry, this results in the condition 
\begin{equation}
    \chi_2<\chi<\pi~.
\end{equation}
Hence, assuming that conditions 1 and 4 are satisfied, then conditions 2 and 3 will be simultaneously satisfied for
\begin{equation}
    \chi_2<\chi<\pi-\chi_2~.
\end{equation}
In the regime where $\alpha\ll 1$, Eq.~\eqref{cotchi} implies $\chi_2\sim \alpha$. 

\begin{figure*}
    \centering
    \includegraphics[width=.8\textwidth]{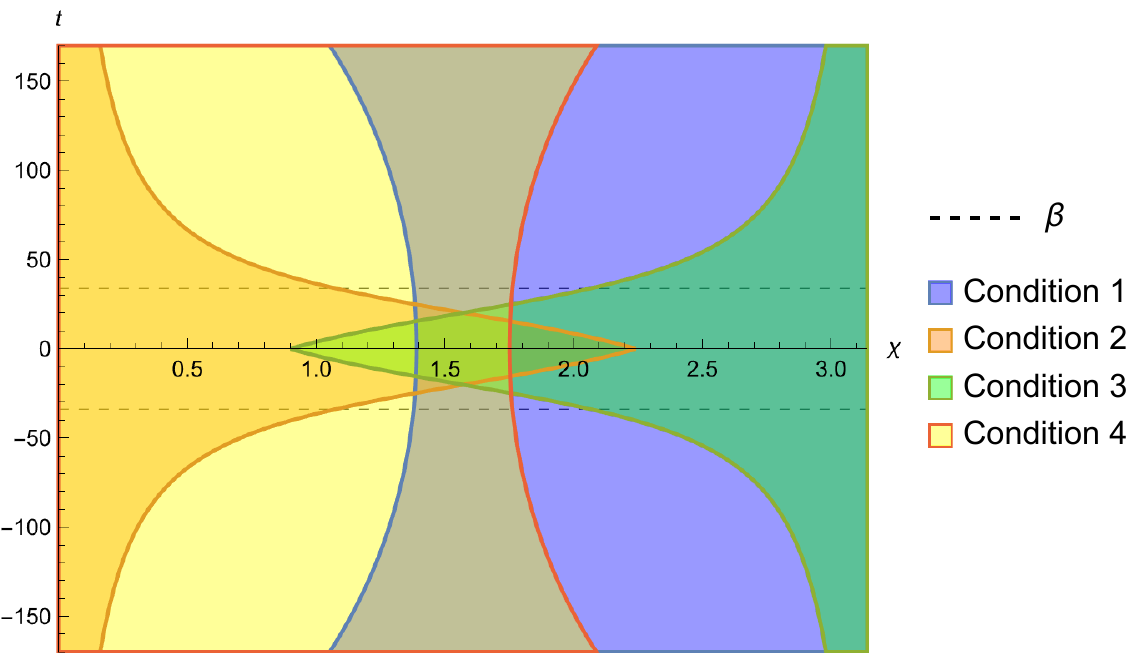}
    \caption{Island conditions in a closed universe with $\Lambda<0$, $t_C \gg t_\Lambda$. Chosen for display is $t_C=25000t_P$, $t_\Lambda=400t_P$. All conditions overlap in a region centered on the equator with temporal width of order $\beta$ around the turnaround time. We verify explicitly that any region at $t=0$ whose only boundary lies in this overlap is an island.}
    \label{fig:plotPKNCC}
\end{figure*}

To summarize, for a subset of the time-symmetric slice to be an island, we require that 
\begin{equation}
    \chi_*<\chi<\pi-\chi_*~,
\end{equation}
where
\begin{equation}
    \chi_*\equiv \max\{\chi_1, \chi_2\}~.
\end{equation}
Importantly, for the conditions 1 and 4 to be simultaneously satisfied, \emph{i.e.}, for the range \eqref{c14range} to be nonempty, we found that Eq.~\eqref{closedc1regime} must be satisfied: $\alpha<\pi/4$. This means that curvature must be dynamically negligible at the turnaround time. 

Near the critical value $\alpha=\pi/4$, $\chi_1$ will be close to $\pi/2$ whereas $\pi/2-\chi_2\sim O(1)$, so $\chi_*=\chi_1$; that is, conditions 1 and 4 are the more stringent. For $\alpha\ll 1$, $\chi_1$ and $\chi_2$ are both of order $\alpha$. A careful analysis keeping O(1) factors shows that $\chi_* =\chi_1$ for all $\alpha$, meaning that conditions 1 and 4 are always more stringent than 2 and 3 at the turnaround time. Thus, to be an island, a subset of the time-symmetric slice must obey
\begin{equation}
    \chi_1<\chi<\pi-\chi_1~.
\end{equation}

Next we examine the full spacetime. The scale factor is
\begin{equation}
    a(t)=t_\Lambda\sqrt{\frac{1}{2}\left(\sqrt{1+\frac{4t_C^2}{t_\Lambda^2}}\cos\left(\frac{2t}{t_\Lambda}\right)-1\right)}~,
\end{equation}
with the turnaround time set to $t=0$. Fig.~\ref{fig:plotPKNCC} shows a check of the four conditions with $t_C \gg t_\Lambda$. As expected from the time-symmetric analysis, the four conditions overlap only in a region centered on the equator with temporal width of order $\beta$ around the turnaround time. (It is worth noting that far from the turnaround time, Conditions 1 and 4 are not always more stringent that conditions 2 and 3.) A check of the full solution with $t_C\sim t_\Lambda$ confirms that there are no islands in that regime.

While we have only verified four necessary conditions, it is easy to check that at the turnaround time, $I(\chi)$ in the range $\chi_1<\chi<\pi-\chi_1$ is indeed an island of a region $R$ of equal size and location on $\Sigma_R$.

Since we found that curvature must be dynamically negligible at the turnaround time for an island $I(\chi)$ to exist, we should be able to make contact with Ref.~\cite{Hartman:2020khs}, which found that on the time-symmetric slice of a flat FRW universe with $\Lambda<0$, any $I(r)$ with proper area radius $r\gtrsim t_\Lambda^{3/2}/l_P^{1/2}$ is an island. Indeed, in a closed universe, the proper area radius of the minimum island at the turnaround time is
\begin{equation}
    r=a_0\sin\chi_*\sim a_0\chi_*\sim (t_C t_\Lambda)^{1/2}\, \frac{t_\Lambda}{t_C}\left(\frac{t_C}{l_P}\right)^{1/2}=\frac{t_\Lambda^{3/2}}{l_P^{1/2}}~.
\end{equation}
As expected the curvature timescale drops out, and we recover the flat FRW result. 

However, there is an important difference: in a flat universe there is no maximum island size. In a closed universe, there is; and it is not the trivial upper bound $\chi=\pi$, because condition 4 becomes violated already for smaller values of $\chi$. For $\chi_R>\chi_1$, at $t=0$, the favored island becomes the entire universe $M$. This is sensible: though curvature has no dynamical effect on the evolution of the universe at the turnaround time, it does affect the kinematics (the topology of space), and the island rule is sensitive to both.

As before, $M$ itself trivially satisfies conditions 1--3, meaning it is a viable island candidate. Let us check when $M$ is in fact an island for some region $R\subset \Sigma_R$. As in the $\Lambda = 0$ case, $R$ must be more than half of $\Sigma_R$ for $R\cup M$ to have less entropy than $R$, \emph{i.e.}, for $M$ to be preferred over the empty set. Now, however, $M$ must also compete with its own subsets. $M$ wins if and only if condition 4 is violated, \emph{i.e.} if
\begin{equation}
    \chi_R>\pi-\chi_1~.
\end{equation}
Since $\chi_1<\pi/2$, this is the only relevant condition.

\subsection{Positive Curvature, Positive Cosmological Constant}
\label{closedPositiveCC}

Let us examine the case where both the cosmological constant and the curvature are positive. As before, we start by finding the scale factor at the turnaround time. In this case, it will be more convenient to work with the Friedmann equation in terms of $\rho_r$, Eq.~\eqref{fried1}. Setting $k=+1$ and $\Dot{a}=0$ implies that the scale factor at the turnaround time is
\begin{equation}
    \label{equation:a0PCPCC}
    a_0 = \left(\frac{8\pi G_N \rho_r}{3}+\frac{\Lambda}{3}\right)^{-1/2}~.
\end{equation}
First let us consider islands that are a proper subset of the closed universe, \emph{i.e.} $I(\chi)$ with $\chi<\pi$. Using the fact that $s\sim\rho_r^{3/4}$ and so $s_c \sim \rho_r^{3/4}a^3$, condition 1 becomes
\begin{equation}
    \frac{V_c(\chi)}{A_c(\chi)}\geq \frac{1}{4\rho_r^{3/4}a_0 l_P^2}~,
\end{equation}
and condition 4 yields the same inequality with ($\chi\rightarrow \pi-\chi$). As before, the fact that $V_c/A_c$ is monotonic in $\chi$ implies that an island candidate can satisfy conditions 1 and 4 simultaneously only if
\begin{equation}
    \label{equation:cond1&4PCPCC}
    \frac{1}{4\rho_r^{3/4}a_0 l_P^2} < \frac{V_c(\pi/2)}{A_c(\pi/2)} = \frac{\pi}{4}~.
\end{equation}

First consider the case where the radiation density dominates at turnaround, $G_N \rho_r \gg\Lambda$. In this regime,
\begin{equation}
    a_0 \sim \frac{1}{\sqrt{G_N \rho_r}} = \frac{1}{\rho_r^{1/2}l_P}~,
\end{equation}
and Eq.~\eqref{equation:cond1&4PCPCC} becomes
\begin{equation}
    \label{equation:cond1&4PCPCClargerho}
    \frac{1}{\rho_r^{1/4}}\lesssim l_P~.
\end{equation}
But the semiclassical regime requires $a_0\gg l_P$ and hence
\begin{equation}
    \frac{1}{\rho_r^{1/4}}\gg l_P~.
\end{equation}
Since these equations are mutually incompatible, conditions 1 and 4 cannot be simultaneously satisfied.

Next, consider the opposite regime where vacuum energy dominates at turnaround, $\Lambda\gg G_N \rho_r$. Then \begin{equation}
    a_0\sim \frac{1}{\sqrt{\Lambda}}\sim t_\Lambda~,
\end{equation}
and Eq.~\eqref{equation:cond1&4PCPCC} becomes
\begin{equation}
    \sqrt{\frac{\Lambda}{\rho_r G_N}}\left(\frac{1}{\sqrt{G_N}\rho_r^{1/4}}\right) \lesssim 1~.
\end{equation}
The first factor is large by assumption, and the second is large since $\rho_r$ cannot approach the Planck density. Hence, this inequality cannot be satisfied for $\Lambda\gg G_N \rho_r$ in the semiclassical regime.
It is easy to verify that the problem persists in the intermediate regime $\Lambda\sim G_N \rho_r$. Thus we have shown (within the parameters of our model) that no proper subset of a time-symmetric slice of a closed universe with positive cosmological constant can be an island.

Next we examine the full spacetime. The scale factor can take one of three forms:
\begin{align}
    \label{equation:aPKPCCexpansion}
   a(t)&=t_\Lambda\sqrt{\frac{1}{2}\left(1-\cosh\left(\frac{2t}{t_\Lambda}\right) +\frac{2t_C}{t_\Lambda}\sinh\left(\frac{2t}{t_\Lambda}\right)\right)}~~~~~~\mbox{(expansion)}\\
    \label{equation:aPKPCCrecollapse}
    a(t)&=t_\Lambda\sqrt{\frac{1}{2}\left(1-\cosh\left(\frac{2t}{t_\Lambda}\right)\sqrt{1-\frac{4t_C^2}{t_\Lambda^2}}\right)}~~~~~~\mbox{(recollapse)}\\
    \label{equation:aPKPCCbounce}
    a(t)&=t_\Lambda\sqrt{\frac{1}{2}\left(1+\frac{1-\xi t_\Lambda^2}{1+\xi t_\Lambda^2} \cosh{\left(\frac{2t}{t_\Lambda}\right)}\right)}~~~~~~\mbox{(bounce)}
\end{align}
where $\xi\equiv 8\pi G_N\rho_r(0)/3$. Eq.~\eqref{equation:aPKPCCexpansion} describes a universe with a big bang at $t=0$ which expands eternally if $t_C/t_\Lambda > 1/2$ or recollapses if $t_C/t_\Lambda <1/2$. Eq.~\eqref{equation:aPKPCCrecollapse} is the same solution as Eq.~\eqref{equation:aPKPCCexpansion} but defined only for $t_C/t_\Lambda <1/2$ with the turnaround time set to $t=0$. Eq.~\eqref{equation:aPKPCCbounce} describes a universe that bounces (the scale factor reaches a minimum) at $t=0$.

\begin{figure*}
    \centering
    \includegraphics[width=.8\textwidth]{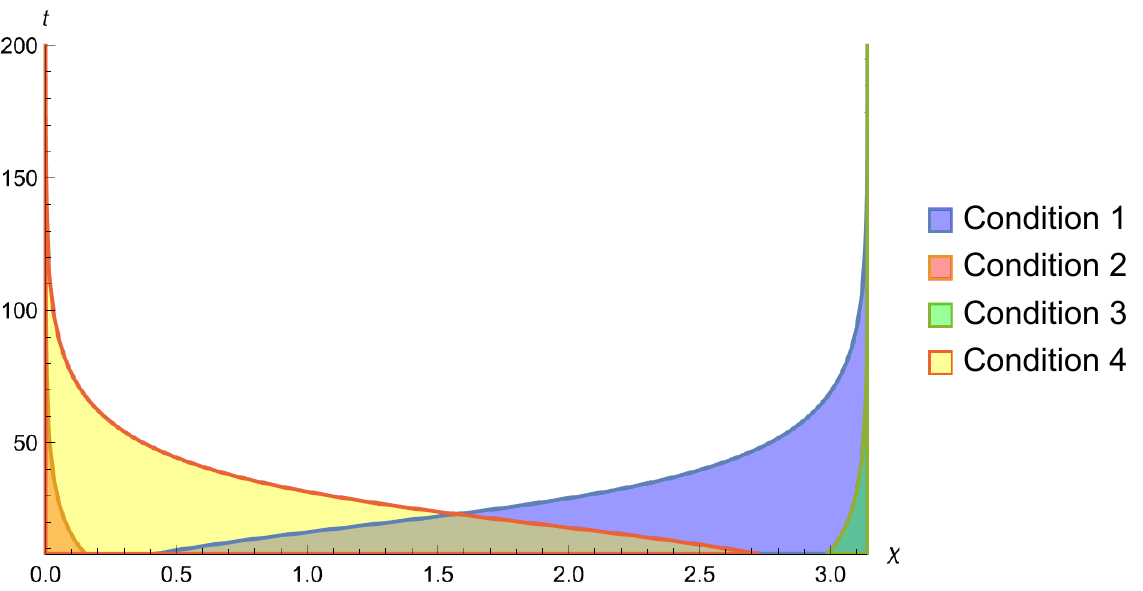}
    \caption{Island conditions in a closed universe with $\Lambda>0$, $t_C \gg t_\Lambda$. Chosen for display is $t_C=1000t_P$, $t_\Lambda=20t_P$. Such a universe expands eternally and thus has no time-symmetric slice. There is no region in which all four conditions are satisfied, meaning there cannot be islands.}
    \label{fig:plotPKPCCexpand}
\end{figure*}

\begin{figure*}
    \centering
    \includegraphics[width=.8\textwidth]{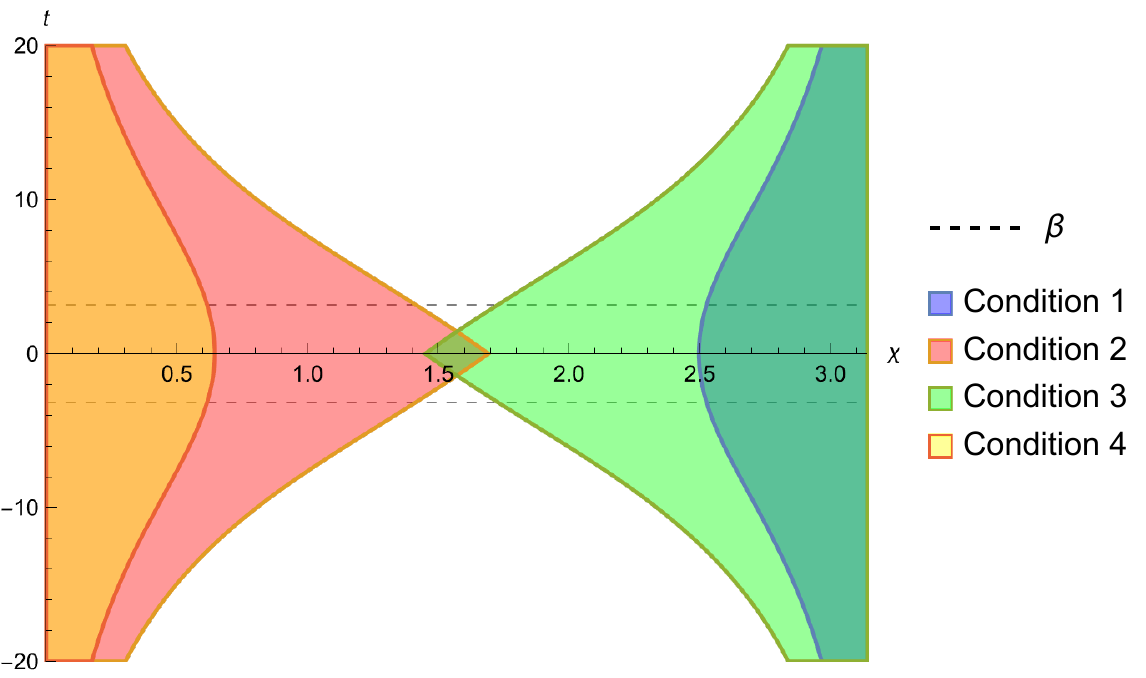}
    \caption{Regions satisfying the four island conditions for a closed universe with $\Lambda>0$ that bounces at $t=0$. The radiation temperature is $\beta^{-1}$ at the turnaround time $t=0$. Chosen for display is $\xi=0.084t_P^{-2}$, $t_\Lambda=10t_P$. There is no region of four-way overlap, so there cannot be islands.}
    \label{fig:plotPKPCCbounce}
\end{figure*}

The expanding solution has no turnaround time, and a check of the four conditions (Fig.~\ref{fig:plotPKPCCexpand}) confirms that there is no region of four-way overlap in its regime ($t_C\gtrsim t_\Lambda$). The recollapsing solution appears qualitatively like the $\Lambda=0$ case and similarly disallows islands. The bounce solution also has no region of four-way overlap (Fig.~\ref{fig:plotPKPCCbounce}).

As in the $\Lambda\leq 0$ cases, the entire closed universe $M$ satisfies all necessary conditions. $M$ will be an island when $R$ is more than half of $\Sigma_R$.
 
\section{Open Universes}
\label{open}

Next we search for islands in universes with negative spatial curvature (open FRW). As before, we start with time-symmetric slices. By Eq.~\eqref{fried2}, none exist for $\Lambda\geq 0$, so we shall take $\Lambda<0$.

The comoving volume and area of a spherical region of coordinate radius $\chi$ are
\begin{gather}\label{vcopen}
    V_c(\chi) = \pi (\sinh(2\chi) - 2\chi)~,\\
    A_c(\chi) = 4\pi \sinh^2(\chi)~.\label{acopen}
\end{gather}
Hence
\begin{equation}\label{frac12}
    \frac{V_c}{A_c}\leq \frac{1}{2}~.
\end{equation}

We begin by ruling out islands in the regime $t_\Lambda/t_C \gtrsim 1$. The scale factor at the turnaround time will be
\begin{equation}
    a\sim t_\Lambda~.
\end{equation}
Using Eqs.~\eqref{equation:s} and \eqref{equation:condition1TS}, condition 1 becomes
\begin{equation}
    \frac{V_c(\chi)}{A_c(\chi)}\gtrsim \left(\frac{t_\Lambda}{t_C}\right)^2 \left(\frac{t_C}{l_P}\right)^{1/2}~.
\end{equation}
The semiclassical regime requires that $t_C/l_P \gg 1$, and we are currently working in the regime $t_\Lambda/t_C\gtrsim 1$, so the r.h.s.\ is large. This conflicts with Eq.~\eqref{frac12}, so condition 1 cannot be satisfied.

Now consider the complementary regime, $t_\Lambda/t_C \ll 1$. The scale factor at the turnaround time is
\begin{equation}
    a_0\sim \sqrt{t_\Lambda t_C}~,
\end{equation}
and condition 1 becomes
\begin{equation}\label{c1open}
    \frac{V_c (\chi)}{A_c (\chi)} \geq \gamma \sim \left(\frac{t_\Lambda}{t_C}\right)\left(\frac{t_C}{l_P}\right)^{1/2}~.
\end{equation}
By Eq.~\eqref{frac12}, this condition can be satisfied only if $\gamma<1/2$. This implies
\begin{equation}
    \left(\frac{t_\Lambda}{t_C}\right)\lesssim\left(\frac{l_P}{t_C}\right)^{1/2}\ll 1~,
\end{equation}
where the second inequality is required for a semiclassical solution. Solving \eqref{c1open} as an equality yields a critical value $\chi_3$ such that condition 1 is satisfied for all $\chi > \chi_3$. Therefore, condition 1 can be satisfied for a spherical island candidate with large enough $\chi$ at the turnaround time if $t_\Lambda$ is early enough.
\begin{figure*}
    \centering
    \includegraphics[width=.8\textwidth]{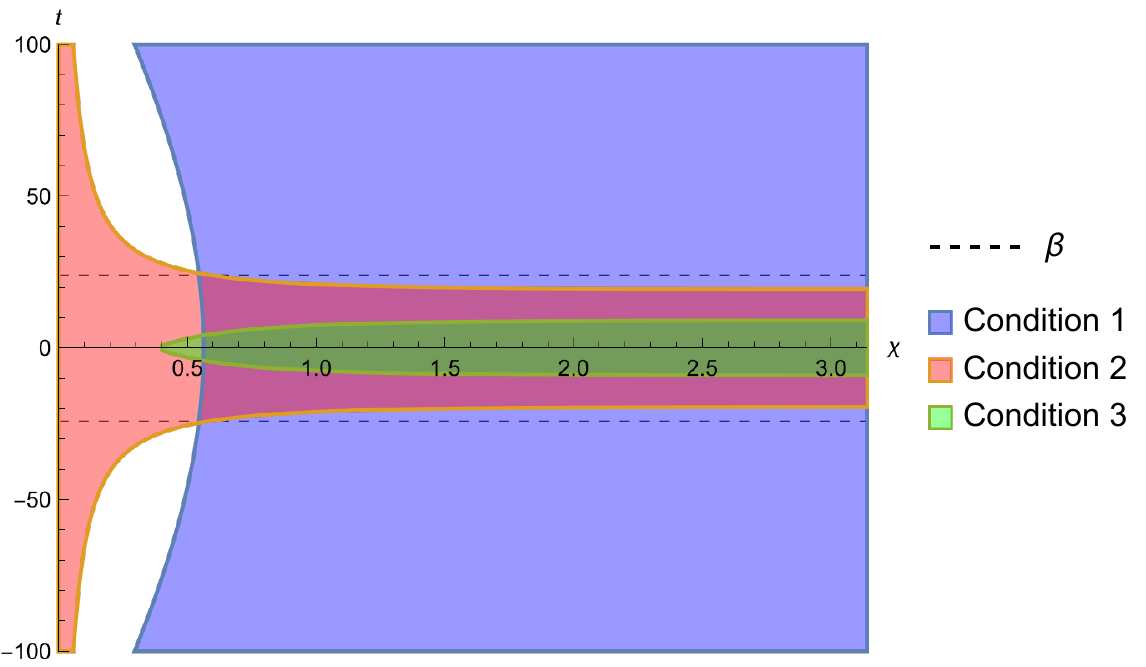}
    \caption{Island conditions in an open universe with $\Lambda<0$, $t_C\gg t_\Lambda$. Chosen for display is $t_C=77500t_P$, $t_\Lambda=200t_P$. All conditions overlap only in a region with temporal width of order $\beta$ around the turnaround time for large enough $\chi$. We verify explicitly that any region at $t=0$ whose only boundary lies in this overlap is an island.}
    \label{fig:plotNKNCC}
\end{figure*}

Since it is possible to satisfy condition 1, we move on to conditions 2 and 3, equations \eqref{equation:condition2TS} and \eqref{equation:condition3TS}. (Condition 4 only applies to subsets of closed universes.) Applying the quantum-normalcy conditions to $I$ and its complement $G$ yields, respectively,
\begin{gather}
    \coth \chi \gtrsim - \gamma^{-1}~,~~~~\mathrm{(Condition~2)}\\
    \coth\chi \lesssim \gamma^{-1}~.~~~~\mathrm{(Condition~3)}
    \label{c3open}
\end{gather}
Condition 2 is satisfied for any $\chi$. Condition 3 can only be satisfied if $\gamma <1$, but we already obtained the stronger restriction $\gamma<1/2$ from condition 1.

Solving \eqref{c3open} as an equality yields a critical radius $\chi_4$, such that all $\chi>\chi_4$ satisfy condition 3. Thus, to be an island, the region must satisfy
\begin{equation}
    \chi^*<\chi<\pi-\chi^*~,
\end{equation}
where
\begin{equation}
    \chi^*\equiv \max\{\chi_3, \chi_4\}~.
\end{equation}
As in Sec.~\ref{closedNegativeCC}, a careful analysis keeping $O(1)$ factors indicates that $\chi_3>\chi_4$ for all $\gamma<1/2$ at turnaround. Thus, condition 1 is always more stringent that condition 3 at the turnaround time. 

Having completed our analysis of the time-symmetric slice, we check the full spacetime. The forms of the scale factor are
\begin{align}
    \label{equation:aNK0CC}
    &\mbox{($\Lambda=0$)}~~~~~a(\eta)=t_C \sinh{\eta}\\
    \label{equation:aNKPCC}
    &\mbox{$(\Lambda>0)$}~~~~~a(t)= t_\Lambda\sqrt{\sinh\left(\frac{t}{t_\Lambda}\right)\left(\frac{2t_C}{t_\Lambda}\cosh\left(\frac{t}{t_\Lambda}\right)+\sinh\left(\frac{t}{t_\Lambda}\right)\right)}\\
    \label{equation:aNKNCC}
    &\mbox{$(\Lambda<0)$}~~~~~a(t)=t_\Lambda\sqrt{\frac{1}{2}\left(\sqrt{1+\frac{4t_C^2}{t_\Lambda^2}}\cos\left(\frac{2t}{t_\Lambda}\right)+1\right)}~.
\end{align}
Eqs.~\eqref{equation:aNK0CC} and \eqref{equation:aNKPCC} describe universes that expand eternally and thus have no time symmetric slice. As expected, these two cases disallow islands. Universes described by Eq.~\eqref{equation:aNKNCC} do recollapse, and in the regime $t_C\gg t_\Lambda$ all three conditions overlap only for large enough $\chi$ in a region with width of order $\beta$ around the turnaround time (Fig.~\ref{fig:plotNKNCC}). Checking the regime $t_\Lambda \gtrsim t_C$ confirms that no islands are possible in that case.

To summarize, in an open universe with $\Lambda<0$, spherical regions with $\chi>\chi_3$ satisfy all necessary island conditions if $\gamma<1/2$, where $\gamma$ is given in Eq.~\eqref{c1open}. This corresponds to a universe in which curvature never dominates since $t_\Lambda/t_C \ll 1$; in fact, curvature cannot dominate even on the scale of the minimum island size. It is easy to verify that these candidates are in fact islands at $t=0$, if $R$ is chosen to be the matching region on $\Sigma_R$.

\paragraph{Acknowledgements} We would like to thank C.~Murdia, T.~Rudelius, and Y.~Nomura for helpful discussions and comments. This work was supported in part by the Berkeley Center for Theoretical Physics; by the Department of Energy, Office of Science, Office of High Energy Physics under QuantISED Award DE-SC0019380 and under contract DE-AC02-05CH11231; and by the National Science Foundation under Award Number 2112880. EW is supported by the Berkeley Connect fellowship.

\bibliographystyle{JHEP}
\bibliography{IINFC}
\end{document}